\documentclass[aps,prl,twocolumn,showpacs,amsmath,amssymb]{revtex4}
\usepackage{epsfig}
\usepackage{dcolumn}
\usepackage{bm}

\begin{document}

\title{Spin-charge separation and Kondo effect in transport
through a 1D Mott-Hubbard insulator. }
\author{V.V. Ponomarenko}
\affiliation{Center of Physics, University of Minho, Campus
Gualtar, 4710-057 Braga, Portugal}

\date{\today}

\begin{abstract}
We study low energy spin and charge transport through a 1D
Mott-Hubbard insulator of finite length $L$ attached to Fermi
liquid reservoirs characterized by different chemical potentials
for electrons of opposite spin polarizations as it happens in
quantum spin Hall insulators. We calculate the average currents
(charge and spin) and their correlators and demonstrate how a
transition induced by the reservoirs to the low energy Fermi
liquid regime results in breakdown of the spin-charge separation,
which is visible in the presence of the spin dependent voltages
and a weak one electron scattering in the system. These
calculations are carried out under assumption that the Hubbard gap
$2M$ is large enough: $M > T_L \equiv v_c/L$ ($v_c$: charge
velocity in the wire) and the scattering rate $\Gamma_s \ll T_L$.
Relation of these results to Kondo dot transport in the Toulouse
limit is also clarified.
\end{abstract}

\pacs{71.10.Pm, 72.25.Mk, 73.40.Rw, 85.75.-d}

\maketitle

Spin-charge separation in a 1D Mott-Hubbard insulator (MHI) known
from the exact solution of the 1D Hubbard model \cite{Lieb} at
half-filling and confirmed in experiments with quasi 1D materials
\cite{Kim,Neudert} can lead to an unusual effect if the insulator
is used for transport between Fermi liquid (FL) reservoirs filled
with electrons of opposite spin polarization up to different
chemical potentials. Different spin-dependent chemical potentials
can arise, in particular, in 2D quantum spin Hall insulators
(QSHIs) \cite{QSH}, where transport is carried by pairs of edge
states of up and down spin polarized electrons moving in opposite
directions. Since in MHI the charge transfer realized by holons
carrying one-electron charge $e$ ($e=\hbar=1$ below) is suppressed
by a Hubbard gap $2M$, but spin carrying excitations, spinons,
move freely, the perfect MHI serves as an ideal spin current
transmitter. Any small difference in the spin dependent voltages
$V_\sigma, \sigma=\pm$ reverses direction of one of the spin
polarized currents in spite of large average difference between
the source and drain chemical potentials. In the case of the QSHI
reservoir the spin polarized tunneling currents can be found by
measuring edge currents to the right and left from the juncture
with the MHI wire along the reservoir boundary. If there is only a
single pair of edge states in the reservoir and the spin dependent
voltages are applied anti-symmetrically, the ideal spin current
transmittance through the MHI makes both these edge currents and,
hence, conductance along the reservoir boundary zero, since the
difference in the currents of the outgoing spin polarized edges is
opposite to the difference of the incoming currents. This simple
picture of the transport can be spoilt, however, by one electron
processes which affect spin-charge separation in the MHI.

In this paper we consider effect of the FL reservoirs on low
energy spin-charge separation in the MHI of finite length $L$: $M
> T_L \equiv v_c/L$ ($v_c$: charge velocity in the wire) in the
presence of one-electron impurity backscattering of low rate
$\Gamma_s \ll M$. For low energy ($<T_L$) the problem has been
mapped \cite{1} through a Duality Transform onto the model of a
point scatterer with pseudospin imbedded in TLL and solved by
fermionization. This model is also related to the Toulouse limit
in Kondo dot transport \cite{Toulouse} as discussed below. Its
solution shows a Kondo-like resonance of conductance below the
crossover energy $\Gamma=\Gamma_s+\Gamma_c$, which remains
exponentially small ($\Gamma_c\approx \sqrt{T_L M}
\exp\{-2M/T_L\}$ at half filling) in the absence of impurities due
to rare tunneling of the condensate phase introduced by the
reservoirs, but can grow up to $\sim M^2/E_F$ ($E_F$ is Fermi
energy in the wire) in the range of applicability of this model as
the backscattering increases. We impose spin dependent voltages in
this model to examine the spin-charge separation by calculating
the average charge and spin currents and their zero frequency
correlators and cross-correlator at temperature $T$. Each average
current depends only on either charge $V_c=\sum_\sigma V_\sigma/2$
or spin $V_s=\sum_\sigma \sigma V_\sigma/2$ voltage combination,
respectively. At the voltages or temperature above the crossover
both currents display a very different behavior only weakly
affected by the one electron scattering. In particular, the spin
voltage $|V_s|>\Gamma$ indeed reverses direction of one of the
spin polarized currents even for $|V_c|\gg |V_s|$. The currents
cross-correlator asymptotically vanishes with increase of $\min
|V_{c,s}|/\Gamma$ or $T/\Gamma$, while both diagonal correlators
remain finite. This shows that the spin-charge separation in
transport through the MHI occurs already in the Kondo regime at
the energies less than $T_L(<M)$. Below the crossover the spin
polarized combinations of the average currents and their
correlators approach in the linear voltage limit the correspondent
decoupled one-electron expressions \cite{noise} specified by the
spin dependent voltages and a single parameter of the one-electron
transmission. The latter is equal to $1/(1+\Gamma_s/\Gamma_c)$ due
to exponential enhancement of the one-electron reflection
$\Gamma_s /\Gamma_c $ by the charge condensate tunneling.

Transport through a one-channel wire adiabatically attached to the
left and right FL reservoirs is modeled by a 1D system of
electrons, whose pairwise interaction is local and switched on by
function $\varphi(x)= \theta (x) \theta (L-x)$ inside the wire of
length $L$. Applying bosonization we can describe the charge and
spin density fluctuations $\rho_{b}(x,t)=e_b \partial_x
\phi_{b}(x,t)/(\sqrt{2} \pi), \ b=c,s$, respectively, with (charge
and spin) bosonic fields $\phi_{c,s}$. Without impurities their
Lagrangian symmetrical under the spin rotation reads
\begin{eqnarray}
{\cal L}\!\!&=&\!\!\int dx \sum_{b=c,s} \bigl[ \frac{v_b(x)}{2g_b(x)} \{
\left( \frac{\partial_t \phi_b(t,x) }{v_b \sqrt{4 \pi}}
\right)^2 - \left(\frac{\partial_x \phi_b(t,x)}{\sqrt{4 \pi}}
\right)^2 \}
\nonumber\\
&-&\!\! E_F^2 U_b \varphi(x) \cos(2 \mu_b x/v_b +
\sqrt{2} \phi_b(t,x))/ (\pi v_F) \bigr] \ , \label{1}
\end{eqnarray}
where $v_F(E_F)$ denotes the Fermi velocity(energy) in the
channel. The parameter $\mu_c \equiv \mu$ varies the chemical
potential inside the wire from its zero value at half-filling and
$\mu_s=0$. The constants of the forward scattering differ inside
the wire $g_b(x)=g_b$ for $x \in [0,L]$ from those in the leads
$g_b(x) =1$, and an Umklapp scattering (backscattering) of the
strength $U_c (U_s)$ is introduced inside the wire. The velocities
$v_{c,s}(x)$ change from $v_F$ outside the wire to some constants
$v_{c,s}$ inside it. We can eliminate them rescaling the spacial
coordinate $x_{old}$ in the charge and spin Lagrangians of
(\ref{1}) into $x_{new}\equiv \int^{x_{old}}_0 dy/v_{c,s}(y)$. As
a result, the new coordinate will have an inverse energy dimension
and the length of the wire becomes different for the charge mode
$L \to 1/T_L$ and spin mode $L \to 1/T_L'$. Applying
renormalization-group results of the uniform sin-Gordon model
\cite{solyom} at energies larger than $T_L$ or $T_L'$ we come to
renormalized values of the parameters in (\ref{1}). For the
repulsive interaction when initially $g_s>1>g_c$, the constant
$U_s$ of backscattering flows to zero and $g_s$ to 1, bringing the
spin mode into the regime of the free TLL . The constant $U_c$ of
Umklapp process increases, reaching $v_F/v_c$ at the energy
cut-off corresponding to the mass of the soliton $M$ if the
chemical potential $\mu$ is less than $M$.  Meanwhile, $g_c$ flows
to its free fermion value $g_c=1/2$.

The spin-charge separation in Lagrangian (\ref{1}) can be broken
by additional one-electron scattering which entangles the spin and
charge modes. We account for such a process by including a weak
backscattering on a point impurity potential inside the wire
$0<x_0<L$:
\begin{equation}
{\cal L}_{p}\!= - \frac{2E_F V_{imp}}{\pi} \cos(\frac{\phi_c(t,x_c)}
{\sqrt{2}} + \varphi_0 ) \cos(\frac{\phi_s(t,x_s)}{ \sqrt{2}})
\ , \label{2}
\end{equation}
where $x_{c,s}=x_0/v_{c,s}$, $\varphi_0\equiv \varphi + \mu x_c$
incorporates a phase of the scatterer $\varphi$. The amplitude
$V_{imp}$ of the impurity potential determines transmission
coefficient as $1/(1+V_{imp}^2)$.

\emph{Low energy model} - An effective model for energies lower
than some cut-off $D'$ specified below has been derived \cite{FC}
from the expression for the partition function ${\cal Z}$
associated to the combined Lagrangian (\ref{1}) and (\ref{2})
following Schmid \cite{schmid}. Without impurities the spin and
charge modes are decoupled. After integrating out $\phi_c$ in the
reservoirs the charge mode contribution into ${\cal Z}$ describes
rare tunneling between neighbor degenerate vacua of the massive
charge mode in the wire characterized by the quantized values of
$\sqrt{2} \phi_c(\tau,x)+ 2 \mu x = 2 \pi m $, $m$ is integer.
Variation of $m$ by $a=\pm 1$ relates to passage of a
(anti)soliton through the wire ((anti)-instanton in imaginary time
$\tau$). The tunneling amplitude may be found as $P e^{-s_0/T_L},
s_0=\sqrt{M^2-\mu^2}, (\mu \ll M)$ by mapping \cite{1} onto a free
fermionic model or instanton techniques \cite{FC}. The latter also
evaluates the pre-factor $P=C \times \sqrt{D'}
(s_0^3T_L/M^2)^{1/4}$ with the constant $C$ of the order of 1. The
parameter $D'$ is a high-energy cut-off to the long-time
asymptotics of the kink-kink interaction: $F(\tau)=
\ln\{\sqrt{\tau^2 + 1/ D'^2}\}$ created by the reservoirs. It
varies with $\mu$ and was estimated from the time scale of the
instanton as $D' \simeq \sqrt{M T_L}$ at $\mu=0$ and $ D' \simeq
(M/\mu) T_L$ if $\mu > T_L$. A crucial modification to this
consideration produced by the impurity under the assumption $E_F
V_{imp} \ll M$ ensues from the shift of the $m$-vacuum. Since it
is equal to $(-1)^m E_F A \cos(\phi_s(\tau,x_s)/\sqrt{2}),
\ A=2 V_{imp} \cos \varphi /\pi $ the neighbor vacua become
non-degenerate. This can be accounted for by an auxiliary
pseudospin variable with the correspondent Pauli matrix
$\sigma_3$. The energy splitting becomes an operator
$\hat{\sigma}_3 A E_F \cos(\phi_s(\tau,x_0)/\sqrt{2})$
acting on the pseudospin, and every (anti-)instanton tunneling
reverses the $\sigma_3$-value with the Pauli matrix
$\hat{\sigma}_1$. The interaction $F(\tau)$ coincides with the
pair correlator of some bosonic field $\theta_c$, whose evolution
is ruled by the free TLL Lagrangian ${\cal L}_0[\theta_c]$ (
uniform Lagrangian (\ref{1}) with no interaction). Then, by
ascribing factors $exp(\mp i \theta_c(\tau_j,0)/\sqrt{2})$ to the
(anti-)instanton at the moment $\tau_j$ the functional integral
for the partition function is reduced \cite{1} to a standard
Hamiltonian form ${\cal Z}=cst \times Tr\{ e^{-{\cal H}/T}\}$ with
%
\begin{eqnarray}
{\cal H}={\cal H}_0[\phi_s(x)]\!\!&+&\!\!{\cal H}_0[\theta_c(x)]
- A E_F  \hat{\sigma}_3 \cos(\phi_s(x_s)/
\sqrt{2})
 \nonumber\\
\!\!&-&\!\! 2 P e^{-s_0/T_L} \hat{\sigma}_1 \cos(\theta_c(0)/
\sqrt{2}) \ . \label{5}
\end{eqnarray}
%
\noindent
Here $\phi_s(x)$ and $\theta_c(x)$ are Schr\"{o}dinger's bosonic
operators related to the variables $\phi_s(\tau, x)$ and
$\theta_c(\tau,x)$ of the functional integration. The free TLL
Hamiltonian ${\cal H}_0[\phi_s(x)] \ ({\cal H}_0[\theta_c(x)])$ is
a function of the field $\phi_s(x) \ (\theta_c(x))$ and its
conjugated  corresponding to the free TLL action ${\cal
L}_0[\phi_s] ({\cal L}_0[\theta_c])$, respectively. The model
(\ref{5}) is  equivalent to the initial one (\ref{1}) at low
energy. It relates to a Point Scatterer with internal degree of
freedom in TLL and is solved exactly through fermionization.

\emph{Fermionization} - The Pauli matrices can be written as
$\hat{\sigma}_\alpha = (-1)^{\alpha+1}\frac{i}{ 2} \sum_{\beta,
\gamma} \epsilon^{\alpha, \beta, \gamma} \xi_\beta \xi_\gamma$
with Majorana fermions $\xi_{1,2,3}$ and antisymmetrical tensor
$\epsilon: \ \epsilon^{123}=1 $. Since the interaction in
(\ref{5}) is point-like localized and its evolution involves only
the appropriate time-dependent correlators, we can fermionize it
making use of:
\begin{equation}
\psi_c(0)=-i \sqrt{\frac{D'}{2\pi} }\xi_3 e^{i\frac{ \theta_c(0)}
{\sqrt{2}}}, \psi_s(0)=i \sqrt{\frac{E_F}{2 \pi}}\xi_1
e^{i\frac{\phi_s(x_s)}{ \sqrt{2}}} . \label{psi}
\end{equation}
Here $\psi_{c,s}(0)$ is
the $x=0$ value of the charge (spin) fermionic field,
respectively. These fields have linear dispersions taken after the
related bosonic fields  with momentum cut-offs (equal to the
energy ones) $D'$ and  $E_F$, respectively. All states of negative
energies are filled. Substitution of these
fields into (\ref{5}) produces a free-electron Hamiltonian:
\begin{equation}
{\cal H}_{F}=\sum_{a=c,s} \{-i \int dx \psi^+_a
\partial_x \psi_a +\sqrt{\Gamma_a } [\psi^+_a(0)-\psi_a(0)]\xi
\} \ , \label{6}
\end{equation}
where the interaction reduces to tunneling between the
$\psi_{c,s}$ fermions and the Majorana one $ \xi\equiv\xi_2\ $.
Here $\Gamma_s=\frac{2E_F}{\pi} (\cos( \varphi ) V_{imp})^2 $ is
the rate of the one-electron backscattering and the rate of the
instanton tunneling is $\Gamma_c=2 \pi C^2 \sqrt{T_L s_0^3}
e^{-2s_0/T_L}/M$.

Application of voltages $V_\sigma$ between the left and right
reservoirs due to the shift of their chemical potentials, in
general, different for electrons of opposite spin polarizations
$\sigma=\pm$ makes the system non-equilibrium and can be described
with a gauge transformation $\phi_{c,s} \rightarrow \phi_{c,s}-
\sqrt{2}V_{c,s}t$ in the real-time Lagrangian (\ref{1}). Each
instanton tunneling increases the condensate phase inside the wire
and adds charge $\Delta \phi_c/(\sqrt{2} \pi)=1$ to the left
reservoir. The correspondent change in the energy of the system
equal to $V_c$ causes a shift $\theta_c/ \sqrt{2} \rightarrow
\theta_c / \sqrt{2}- V_c t$  in the $\cos$ argument in
Eq.(\ref{5}). Both transformations then can be accounted for with
the shifts of the charge and spin fermion chemical potentials in
Eq. (\ref{psi}) by $V_{c,s}$, respectively. Assuming below that
both voltages are applied antisymmetrically and small enough,
$|V_{c,s}|<T_L<M$, we neglect their effect on the other parameters
in the fermionized Hamiltonian (\ref{5}).

To find  the charge and spin  currents flowing through the channel
we notice that each antiinstanton realized by $\psi_c(0)$
transfers charge $1$ to the right reservoir with no transfer of
spin. Then the charge current operator is $J_c=-i[\int dx \psi_c
\psi^+_c, {\cal H}_{F}]= i \sqrt{\Gamma_c }[\psi^+_c(0)+\psi_c(0)]
\xi\equiv j_c$. On the other hand, free passage of the spin
current $J_s$ through the channel is affected by backscattering
due to the spin field interaction in Hamiltonian  ${\cal H}$ in
Eq. (\ref{5}). This makes the spin current equal to
$J_s=V_s/\pi-j_s$, where the backscattered spin current operator
can be found as $j_s=-\sqrt2 \delta {\cal H}/\delta \phi_s =i
\sqrt{\Gamma_s }[\psi^+_s(0)+\psi_s(0)] \xi$.

A crucial feature of the Hamiltonian ${\cal H}_F$ in Eq. (\ref{6})
is that its interaction and the currents it creates contain two
different Majorana components $\eta_{a,\mp}(0)$ of each fermionic
field $\psi_a$, respectively. These components are defined by
$\psi_a(0,t)=(\eta_{a,+}-i \eta_{a,-})/\sqrt{2}$. Being orthogonal
at the same time both the $\eta_{a,\pm}(t)$ components lose this
property if taken at different times due to the applied voltages.
Still, in Keldysh technique we need to use in non-equilibrium
calculations both retarded and advanced cross-diagonal Green
functions $g^{R,A}_{a,+,-}(t)$ of free Majorana fermions vanish.
Only the cross-diagonal Green functions
$g^{>,<}_{a,+,-}(t)=-g^{>,<}_{a,-,+}(t)$ are non-zero and equal to
$g^{>,<}_{a,+,-}(\omega)=[f((\omega-V_a)/T) -f((\omega+V_a)/T)]/2$
in the frequency representation, where $f$ is the
Fermi-distribution function. Then, from the Dyson equation the
total cross-diagonal Green function
\begin{equation}
G^{>,<}_{a+,X}=i\sqrt{2\Gamma_a}
(g^{>,<}_{a+,a-}G^A_{\xi,X}+g^R_{a+,a-}G^{>,<}_{\xi,X}) \
\label{Ggamma}
\end{equation}
reduces to the first product on the right-hand side, if the index
$X$ denotes $\xi$ as the second field, or vanishes at all, if
$X=b+$. The diagonal total Green function $G^{>}_{a+,a+}$
coincides with the free one.

\emph{Average currents} - As follows from  Eq. (\ref{Ggamma}) the
average of the operator $<j_a>$ can be written as
$<j_a>=-i2\Gamma_a\int d \omega
g^>_{a,+,-}(\omega)G^A_\xi(\omega)/(2\pi)$ and depends only on the
corespondent voltage $V_a$, since the advanced Green function
$G^A_\xi$ does not contain information about voltages. It is
related to the free Green function $g^{A(R)}_\xi=2/(\omega \mp
i0)$ through the correspondent Dyson equation with the self-energy
$\Sigma^{A(R)}_\xi=\pm i \Gamma$. Substitution of the expressions
for both Green functions results in
\begin{equation}
<\!j_a\!>=\!\frac{2 \Gamma_a \Gamma }{\pi}\!\!\int\! d\omega
\frac{f(\frac{\omega-V_a}{ T})-f(\frac{\omega +V_a}{ T})}
{\omega^2+4\Gamma^2} \, ,
\label{9}
\end{equation}
The average charge and spin currents $\bar{J}_{c,s}$ are equal to
$<\!j_c\!>$ and $V_s/\pi -\! <\!j_s\!>$, respectively. At low
temperature $T<\Gamma$ the integral in Eq. (\ref{9}) converges to
$<j_a>=(2 \Gamma_a / \pi ) \arctan (V_a/(2\Gamma))$. Then the
average spin polarized currents $\bar{J}_\sigma=(\bar{J}_{c}+
\sigma \bar{J}_{s})/2$ below the crossover $\Gamma$ approach the
one-electron expressions $\bar{J}_\sigma=D V_\sigma/(2 \pi)$,
where the spin independent transmittance $ D=\frac{\Gamma_c}{
\Gamma}$ demonstrates renormalization of the initial amplitude
$V_{imp}$ in Eq. (\ref{2}) into $\sqrt{\Gamma_s/\Gamma_c}$ by the
interaction inside the MHI. Above the crossover the charge current
saturates at $\Gamma_c$ while the spin current grows up as
$V_s/\pi-\Gamma_s$. Therefore, for  $|V_c|\gg |V_s|$ one of the
conductances $\bar{J}_\sigma/V_\sigma$ becomes negative as $|V_s|
\agt \pi \Gamma$ suggesting emergence of the spin-charge
separation. Similarly, the separation may be expected at $T \gg
\Gamma$. Indeed, the linear bias charge conductance is $ G_c=
\Gamma_c\psi'\left(1 /2+\Gamma/(\pi T)\right)/(\pi^2 T)$, where
$\psi'(x)$ is the derivative of the di-gamma function,
$\psi'(1/2)=\pi^2/2$. The high temperature asymptotics of both
conductances  $G_c=\Gamma_c /(2T)$ and $G_s= 1 / \pi - \Gamma_s
/(2T)$ are defined by different $\Gamma$-parameters and
independent of each other, and the condition $1 \gg
|V_s/V_c|=G_c/G_s$ on the current reversing voltages is satisfied.
We further examine correlators between the charge and spin
currents.

\emph{Currents correlators} -  The zero-frequency current
correlators $\delta^2J_{ab}$ are related to the current operators
correlators $\delta^2 j_{ab}\!=\!\int dt \exp[-i \omega t] <\!j_a(t)
j_b(0)\!>, \omega \to 0$ in the following way $\delta^2 J_{ab}=\pm
\delta^2 j_{ab}$, where $\pm$ stands for diagonal and
cross-correltors, respectively. The diagonal spin current
correlator at finite temperature also includes\cite{PRB}
additional  terms $\delta^2J_{ss}\!=\! 2T/\pi - 4T
\partial_{V_s}\!<\!j_s\!> + \delta^2j_{ss}$. Appearance of these
terms may be easily understood recalling that the
fluctuation-dissipation theorem claims non-zero current
fluctuations even in the absence of backscattering at non-zero
temperature.

Substituting expressions for the current operators $j_{a,b}$ into
their correlator and then splitting the correlator into pair-wise
correlators of the Majorana fields by applying Wick's theorem we
find that
\begin{eqnarray}
\delta^2 j_{ab}=&&\!\!\!\! - 2 \sqrt{\Gamma_a \Gamma_b}\!\! \int\! \frac{d
\omega}{ 2 \pi} [G^>_{a+,\xi}(\omega)G^<_{a+,\xi}(\omega)
\nonumber \\
+&&\!\!\!\!G^>_{a+b+}(-\omega) G^>_\xi(\omega)]\equiv
I^{(1)}_{ab}+\delta_{a,b}I^{(2)}_{a} \  \label{deltaJ}
\end{eqnarray}
Then the cross-correlator of the two currents equal to $-
I^{(1)}_{cs}$ in Eq. (\ref{deltaJ}) follows from Eq.
(\ref{Ggamma}) as
\begin{eqnarray}
\delta^2J_{cs}=-\frac{4 \Gamma_c \Gamma_s}{ 2 \pi} \int d \omega
[G^A_\xi(\omega)]^2g^>_{c+-}(\omega)g^<_{s+-}(\omega)
\nonumber \\
 =-\frac{4 \Gamma_c \Gamma_s}{ \pi} \int d \omega \frac{\prod_{a=c,s}
\left[f \left(\frac{\omega-V_a}{ T}\right) - f \left(\frac{\omega+V_a
}{ T}\right) \right]}{(\omega-2i\Gamma)^2} \ . \label{11}
\end{eqnarray}
\noindent
It is vanishing with
increase of both voltages or temperature under the integral
 in Eq. (\ref{11}) due to the analytical structure of the advanced Green
function. In particular, in
the limit of $|V_{c,s}|/T \gg 1$ it takes the form
\begin{equation}
\delta^2J_{cs}=\frac{4\Gamma_c \Gamma_s}{ \pi}
\frac{\mbox{sign}(V_cV_s) V} {[4\Gamma^2+ V^2]}, \   V=
\min_{a=c,s} |V_a|\, , \label{csTzero}
\end{equation}
which approaches zero $\propto \Gamma_c \Gamma_s/V$ as $V$ becomes
much larger than $2 \Gamma$. In general,  the integral in Eq.
(\ref{11}) is expressed in terms of derivatives of the di-gamma
function. This expression shows that at high temperature $T\gg
2\Gamma$ the cross-correlator is vanishing as $
\delta^2J_{cs}=|\psi^{(2)}(1/2)+\psi^{(4)}(1/2)| \Gamma_c \Gamma_s
V_c V_s/(12 \pi^5 T^3)$ if $1 \gg \max|V_{c,s}|/T$.

The diagonal correlators of the charge and spin currents also
include the second term in Eq. (\ref{deltaJ}). To find it we
notice that the Green function $G^>_\xi$ of the single Majorana
operator reduces to $G^>_\xi=G^R_\xi\Sigma^>_\xi G^A_\xi$ with the
self energy $\Sigma^>_\xi=-i\sum_{a,\pm} \Gamma_a f((-\omega \pm
V_a)/T)$ and $g^>_{a++}(-\omega)=-i\sum_\pm f((\omega \pm
V_a)/T)$. Substitution of these expressions into Eq.
(\ref{deltaJ}) gives us the diagonal zero-frequency correlator of
the charge current in the following form
\begin{eqnarray}
\delta^2J_{cc}=\frac{2 \Gamma_c}{\pi}\int d\omega
\frac{\left[f(\frac{\omega+V_c}{ T})+f(\frac{\omega -V_c}{
T})\right]}{\omega^2+4\Gamma^2}
\nonumber \\
\times \sum_{a=c,s}\Gamma_a \left[f(\frac{-\omega+V_a}{
T})+f(\frac{-\omega -V_a}{ T})\right] + I^{(1)}_{cc} \  \label{12}
\end{eqnarray}
and the similar expression for correlator of the spin backscattered
current $\delta^2j_{ss}$ by exchanging the "c" and "s"-indexes.
At zero temperature the latter correlator coincides with
$\delta^2 J_{ss}$ and hence both shot noises of the charge and
spin current ($a=c,s$) rise from Eq. (\ref{12}) as
follows:
%
\begin{eqnarray}
\delta^2J_{aa}&&\!\!\!\!= \left( \frac{ 2\Gamma_a^2 }{\Gamma
\pi} \arctan\left(\frac{|V_a|}{ 2 \Gamma }\right) +
\frac{2  \Gamma_c \Gamma_s}{\pi \Gamma} \right. \nonumber \\
\times&&\!\!\!\!\!  \left. \arctan\left(\frac{\max\{|V_b|\}_{b=c,s}}
{ 2 \Gamma }\right) -\frac{ 4\Gamma_a^2  |V_a|}{ \pi
(V_a^2+4 \Gamma^2)} \right) \ . \label{13}
\end{eqnarray}
%
\noindent
In the low voltage limit both
expressions coincide
$\delta^2J_{cc}=\delta^2J_{ss}=(1/ \pi)
D(1-D)\max\{|V_b|\}_{b=c,s}$.  Combined with the low voltage
cross-correlator in Eq. (\ref{csTzero}) they give
shot noise of the two spin polarized currents  \cite{noise}
in the one-electron form $\delta^2J_{\sigma\sigma'}=(\delta_{\sigma\sigma'}/ \pi)
D(1-D)|V_\sigma|$.

As
both voltages increase above the crossover the shot noise in the
two currents grows and saturates at two different values
$\delta^2J_{aa}\!=\!\Gamma_a$. Together with demonstrated before
suppression of the cross-correlator of the currents this confirms
the spin-charge separation above the crossover but yet below $T_L$.
Similarly, the charge current noise
$\delta^2J_{cc}=\Gamma_c[1+2 \psi^{(2)}(1/2)\Gamma/(\pi^3
T)]$ remains finite at high-temperature,
while the spin noise grows linearly $\delta^2J_{ss}=2T/\pi-\Gamma_s=2TG_s$.

\emph{Toulouse limit in Kondo dot transport} -  This model has
been written \cite{Toulouse} as a formal generalization of the
physical model of Kondo dot transport. It describes tunneling
between two branches, $a=r,l$, of 1D chiral fermions $\psi_{a,\sigma}(x)$
carrying spin $\sigma$
and propagating in the right and left reservoirs, respectively,
with the tunneling Hamiltonian
\begin{equation}
{\cal H}_{T}=\sum_{a,b=r,l} \sum_{\sigma,\sigma',\gamma}
J^{a,b}_\gamma \hat{\tau}_\gamma
\psi^+_{a,\sigma}(0)\sigma^\gamma_{\sigma,\sigma'}
\psi_{b,\sigma'}(0)
 \ , \label{HT}
\end{equation}
where $\hat{\tau}_\gamma$ are the Pauli matrices for the Kondo dot
spin. The choice of the parameters $J^{a,a}_z=2\pi, J^{r,l}_z=0,
J^{a,b}_{x,y}=J^{a,b}_\perp $ corresponds to the Toulouse limit
model solvable through bosonization and Emery-Kivelson rotation.
Under additional restriction $J^{r,r}_\perp=-J^{l,l}_\perp $ its
Hamiltonian takes the form of ${\cal H}$ in Eq. (\ref{5}) with the
field $\phi_s$ substituted by its dual $\theta_s$, which describes
tunneling of the unit of spin instead of its backscattering, and
further coincides with ${\cal H}_F $ in Eq. (\ref{6}) after the
fermionizaton  with $\Gamma_c=E_F [J^{r,l}_\perp]^2/(8 \pi),
\Gamma_s=E_F [J^{r,r}_\perp]^2/(8 \pi) $. Both operators $j_{c,s}$
now describe tunneling of the charge and the spin. Interchanging
the direct and backscattered spin currents in the above
calculation we apply their results to the Toulouse limit model. In
particular, the low energy linear voltage dependence of the
average charge and spin currents is defined by different
transmittances equal to $D$ and $1-D$, respectively. This feature
can not be accounted for in the one-electron transport model.
Indeed, in this model the spin cross-diagonal transmission and
reflection coefficients become $\pm(D-1/2)$, and one of them would
be negative unless $D=1/2$. This indicates independent tunneling
of spinons and break of the low energy electronic FL description.

In summary, we have shown that in spite of influence of the FL
reservoirs and a weak one-electron scattering the essential
properties of the spin-charge separation in transport through a
MHI of finite length appear if either $T$ or both $|V_{c,s}|$ are
above a crossover energy ($\ll T_L$), below which, however, the
transport becomes one-electron.

\end{document}